# Exploring Room-Temperature Superconductivity in Narrow Energy Gap Semiconductors through Thermally Excited Electrons

Ning Chen[1], Yang Liu[1], and Yang Li[2],
1 School of Materials Science and Engineering, University of Science and Technology Beijing, Beijing, 100083, P.R. China
2 Department of Engineering Science and Materials, University of Puerto Rico, Mayaguez, Puerto Rico 00681-9000, USA

**Abstract:** The impact of thermally activated electrons on superconductivity within the realm of narrow energy gap semiconductors is investigated, unveiling the potential emergence of room-temperature superconductivity without the necessity for extensive doping. This novel mechanism suggests that thermal excitations can influence carrier concentrations significantly, thus influencing condensation and affecting superconductivity properties. In this paper, we explore how the energy gap width and temperature interact to shape carrier concentrations and subsequently impact the manifestation of superconducting behavior.

**Key words**: superconductivity, room temperature superconductor, lead apatite, thermal excitation

## Introduction

Superconductivity, a phenomenon observed primarily in metals, intermetallic compounds, and doped semiconductors, remains elusive in non-doped insulators and semiconductors. The carrier concentration in the latter materials often falls short of the required threshold. This study delves into the potential of thermally excited electrons within narrow energy gap semiconductors to drive the appearance of superconductivity at elevated temperatures, potentially leading to room-temperature superconductivity.

The carrier concentration near the Fermi level plays a critical role in the superconducting behavior of various materials. For instance, insulators possess a substantial energy gap that hinders electron transitions, while metals readily conduct due to partially filled conduction bands. In doped semiconductors, exemplified by high-temperature cuprate superconductors, carrier concentration changes might be influenced by thermally excited electrons. However, their effects on the critical temperature (30~138 K) are limited. In contrast, narrow energy gap semiconductors exhibit sensitivity to temperature-induced variations in carrier concentration—a characteristic that significantly impacts room-temperature superconducting properties.

Considering a hypothetical scenario of room-temperature superconductivity, it becomes crucial to analyze the temperature-dependent behavior of carrier concentrations. As temperature decreases from room-temperature, a rapid decline in initial resistance due to superconductivity might be observed. But the total carrier concentration supplied to superconducting regions decreases with temperature, leading to altered zero-resistance and diamagnetic measurements.

**Mechanism and Analysis**

To analyze the properties of a novel room-temperature superconducting system, temperature becomes a critical factor. Since the critical temperature of a room-temperature superconductor is often much higher than that of conventional types like cuprates, studying the carrier concentration near the Fermi level through thermally influenced electrons at high temperatures is essential, as shown in Figure 1. Furthermore, the critical temperature at room temperature is also four times higher than that of the highest copper oxide system, potentially resulting in a concentration difference of $10^7$ times for a $0.5eV$ energy gap semiconductor, according to the electronic Fermi-Dirac distribution. Hence, studying the influence of high temperature on carrier concentration is necessary. At room temperature, while the band structure remains largely unchanged, the concentration of thermally activated electrons occupying the Fermi level undergoes significant changes. These changes could potentially impact condensation—an essential distinction between room-temperature superconductivity and other superconductor systems. It's evident that all previous superconducting types possess a critical temperature (<138K) that is too low for this issue to be discussed.

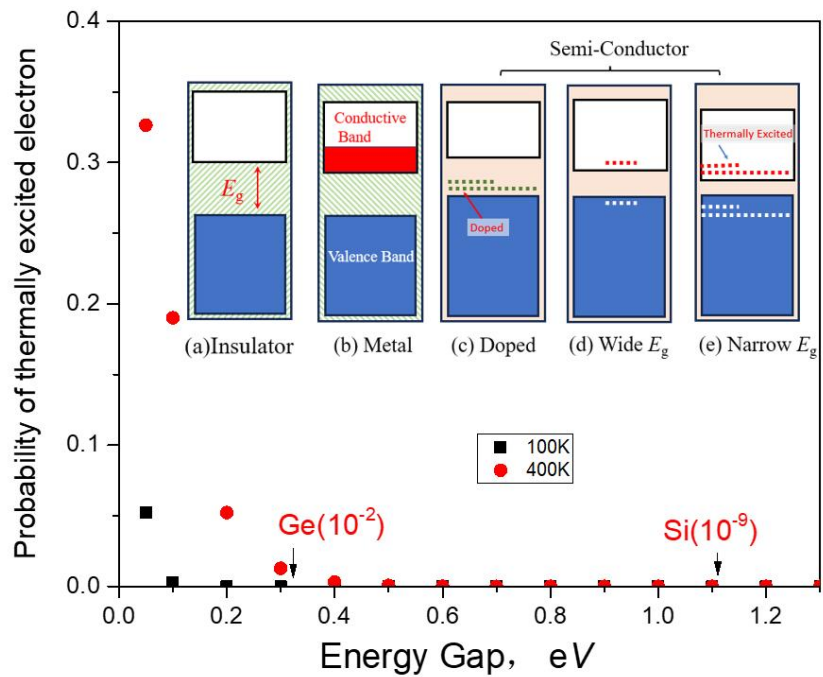

Fig. 1. Compared probability of thermal activation of 100K and 400K as function of energy gap ($E_g$) of different systems (a)-(e), the occurrence of superconductivity at considerably high temperatures of narrow ($E_g$) system is significant as the influence of temperature on the carrier concentration near the Fermi level through thermal excitation.

Fig. 1(a) and 1(b) illustrate insulator and metal systems, respectively. The large energy gap of insulators makes it challenging for electrons to transition to the conductive band even above room temperature. This is because the conductive band remains mostly empty, lacking free electrons for conduction. On the other hand, metals possess a half-filled conductive band, resulting in a considerable number of electrons participating in conduction around the Fermi level at room temperature. Consequently, the count of conductive electrons is minimally affected by temperature changes in metal systems.

Fig. 1(c) and 1(d) correspond to the band structure of high-temperature cuprate superconductors, categorized as doped semiconductor types where carrier concentration is largely determined by doping levels. The change in carriers at the Fermi level induced by thermally excited electrons can be calculated according to the Fermi-Dirac distribution of electrons as a function of energy gap versus temperature[1]. However, this change doesn't significantly alter the critical temperature range of the high-temperature superconducting region (<138K) and can thus be disregarded.

Fig. 1(d) and 1(e) correspond to two intrinsic semiconductors with narrow energy gaps ($E_g$) and no doping. Carrier concentration is notably sensitive at room temperature. In narrow-gap semiconductors, a certain number of carrier concentrations can already occur at room temperature, with the energy gap directly determining the count of thermally excited carriers.

The following section explores whether room-temperature superconducting systems would possess different property characteristics compared to existing superconducting systems. While achieving room temperature superconductivity remains highly challenging due to the conditions required for superconducting pairing intensity, the ultra-high temperature offers a beneficial factor for achieving higher electron concentration at Fermi levels. According to the Fermi-Dirac distribution of electrons, the number of thermally excited electrons in an empty orbit near the Fermi surface is directly proportional to the energy gap width.

In this context, carrier concentration in such a semiconductor is a function of temperature and energy gap. If the energy gap of a semiconductor is sufficiently narrow, for instance, with an energy gap less than $0.2eV$, it's possible to obtain enough carriers through thermal excitation at room-temperature. This might enable the attainment of carrier concentrations comparable to those achievable through doping, meeting the requirements for superconductivity (considering an initial doping concentration of 0.04 per $CuO_2$ plane for a cuprate system). Thus, even without any externally introduced doped carriers, suitable carrier concentrations could be achieved through intrinsic thermal excitation. This underscores that room-temperature superconductivity can be accomplished without doping as long as the energy gap is adequately narrow. It might be a distinctive aspect of the electronic structure of room-temperature superconductivity.

Continuing, we examine the changes in primary superconductivity measurements as temperature decreases. Assuming that superconductivity can already be achieved through thermal excitation in a narrow energy gap ($E_g$) system, we observe that as temperature drops, the zero-resistance and diamagnetic measurements also exhibit distinctive changes due to the temperature-dependent nature of carrier concentration.

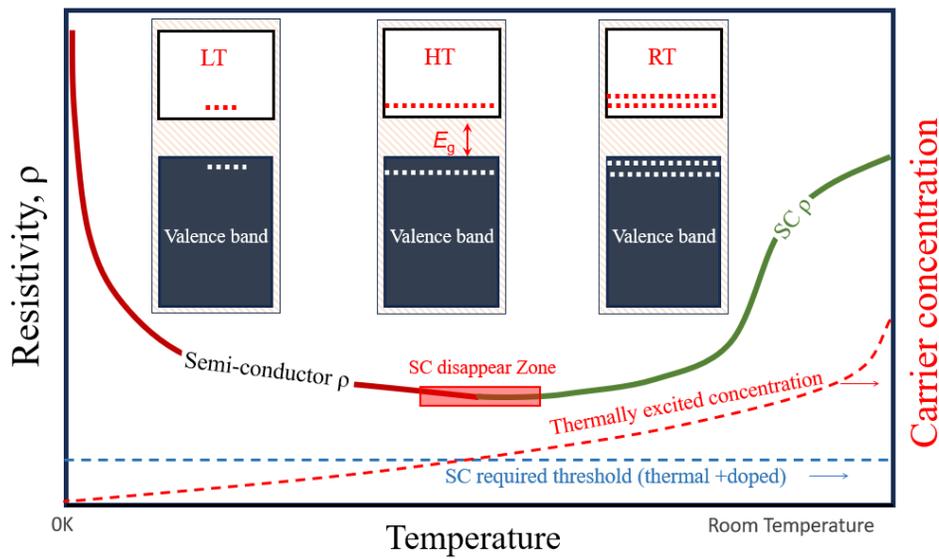

Fig. 2. illustrates the changes in carrier concentration in (a) lower temperature region (LT), (b) high temperature region (HT), and (c) at room temperature (RT). RT superconductivity emerges due to the presence of sufficient intrinsically thermally excited electrons.

Figure 2 demonstrates that as the temperature of the narrow-band semiconductor drops from the critical temperature, the initial resistance experiences a rapid decline. At this juncture, if temperature remains relatively stable, the total number of carriers available for condensation can be disregarded, consequently not affecting the curves of diamagnetic or zero resistance measurements. However, as the temperature range significantly decreases, the total carrier concentration supplied to the superconducting state diminishes significantly. This weakens the superconducting ratio, leading to a relatively gradual decline in zero resistance or a relatively rapid decrease in diamagnetism. Upon reaching a critical concentration point, either superconductivity ceases or the resistance measurement reflects semiconductor characteristics only, or diamagnetism disappears entirely, signifying the vanishing of room-temperature superconductivity.

**Discussion**

Clearly, these unique phenomena or principles governing the peculiar properties of thermally active superconductivity closely resemble some of the current measurement behaviors observed in $Pb_{10-x}Cu_x(PO_4)_6O$[2,3]—weak diamagnetism and the absence of zero resistance, characteristics confirmed by numerous recent experiments. Additionally, some DFT[4,5] calculations suggest that $Pb_{10}(PO_4)_6O$ is a smaller gap semiconductor[6,7]. Notably, through Cu doping substitution, the energy gap can continue to decrease, potentially even disappearing entirely when more than 4 Cu atoms are substituted from 10 Pb atoms. Given that both Pb and Cu possess a +2 valence, doping out carriers becomes improbable, indicating a narrow gap semiconductor[8] also.

Furthermore, in semiconductors with suitable narrow gaps, if insufficient excited electrons exist to sustain superconductivity, a certain concentration of doped carriers becomes essential. However, for ionic crystals, which typically possess an overall band width[9,10] (pairing intensity factor) due to some core orbital couplings, achieving carriers through doping isn't as straightforward as in Mott-Hubbard insulators. In such cases, obtaining carriers via doping is challenging. Consequently, the emergence of ultra-high-temperature-controlled superconductivity primarily relies on the use of an appropriate number of thermally excited electrons. Thus, designing room-temperature superconductors necessitates adjusting the energy gap of the material system to facilitate thermal activation carriers and support superconductivity.

On a different note, if sample superconductivity occurs at room temperature, instability in its physical properties due to thermally excited electrons becomes a common occurrence. Achieving electron pairing intensity[11,12] and sufficient carriers at exceedingly high room temperatures also necessitates meticulous sample preparation. Nonetheless, a sample must exhibit zero resistance and complete diamagnetism in a local region, at least at a certain temperature, to truly qualify as exhibiting superconductivity. These two traits are interdependent, forming two facets of superconductivity, as the supercurrent enabling diamagnetism is unachievable without zero resistance. Should the sample lack uniformity or even possess heterogeneity, the superconducting phase can only manifest as an "island mechanism" or "sausage structure". Consequently, the volume fraction of superconductivity remains low, making it challenging to achieve overall "bulk superconductivity." This is because the limited superconducting phase content cannot form a complete current loop, leading to the absence of zero resistance across the entire sample (thermal excitation and dimensionality exacerbate this problem). Despite this, the absence of "bulk superconductivity" doesn't negate the existence of local superconductivity. Simultaneously, the strength of diamagnetism diminishes significantly due to the low volume fraction of the superconducting phase, likely preventing complete magnetic levitation, yet making diamagnetism relatively easy to determine.

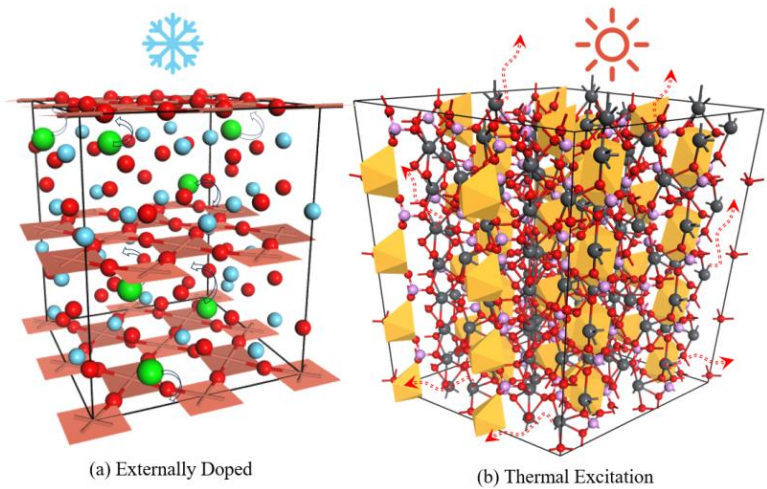

(a) Externally Doped  (b) Thermal Excitation

Fig. 3. Potential for room-temperature superconductivity in narrow energy gap semiconductors through the influence of thermally excited electrons, even without external doped like cuprate.

In conclusion, this study underscores the potential for room-temperature superconductivity in narrow energy gap semiconductors through the influence of thermally excited electrons. The interconnectedness of carrier concentration, energy gap, and temperature unveils a distinctive route toward achieving superconducting behavior. While challenges persist and further research is imperative, the exploration of room-temperature superconductivity holds promise in revealing novel material properties and reshaping our comprehension of superconducting phenomena. At the same time, we must still focus on what key factors (other than carrier factors) cause superconductivity to appear near room temperature.